# Derivation of the Paraxial Ray Equations in Three Conformal Representations

David R. Bergman[1]
Morristown NJ

*Abstract*

Acoustic rays can be described by null geodesics of a pseudo-Riemannian manifold. This allows for the immediate generalization of paraxial ray systems via geodesic deviation. The technique has appeared in the literature for the past decade and has been used by the author in the development of dynamic ray trace codes. The techniques involved are powerful but abstract and frequently unfamiliar in the field of acoustics. This brief paper derives the complete paraxial system for the case of acoustics in the presence of an inhomogeneous isotopic time independent media described by a sound speed profile $c(\vec{r})$. The derivation is performed for three distinct conformal representations of the acoustic metric.

[1] Author's permanent email address: davidrbergman@essc-llc.com

## 1. Introduction

Acoustic rays in a moving fluid are null geodesics of a pseudo-Riemannian (or Lorentzian) manifold. This connection has been documented by researchers in the field of general relativity [1], [2], underwater acoustics [3], and the general treatment of the method of characteristics applied to hyperbolic partial differential equations (PDE) [4]. The author of this paper has made use of the equations of geodesic deviation to develop a paraxial or dynamic ray trace procedure for acoustic rays in a moving fluid medium with background fields, $c$ and $\vec{v}_0$ depending arbitrarily on space and time [5], [6], [7], [8]. Other applications of differential geometry to acoustics are, the use of isometry for discovering conserved currents in the ray equations, reduction of order of the ray equations, and application of conformal symmetry of the null geodesics to transform the paraxial ray system into any representation desired. This last point is related to this paper.

Conformal symmetry implies that there is a continuously infinite number of abstract geometries (acoustic metrics) that will give essentially the same paraxial ray system. The consequences of this are non-trivial from a computational perspective as one choice of conformal representation can lead to a very simple set of equations while another choice can lead to a very complicated system that is both hard to solve and difficult to interpret. There are several ways to represent the paraxial ray system in acoustics but none adhere to a specific conformal representation. In fact a careful inspection of the literature shows that we are frequently dealing with "mixed" representations from the point of view of differential geometry. Each new treatment of the problem leads to slightly different equations which sometimes look like geodesic deviation, sometimes include extra terms involving the first derivative of the deviation vector and sometimes include inhomogeneous terms in the paraxial equation.

We consider a general case of acoustic propagation in the presence of a sound speed profile depending on only three Cartesian coordinates. We further assume the background fluid velocity (wind or current) is zero and that the background density is constant. The full set of equations for describing the paraxial ray system is derived for three specific representations commonly used in the literature. These results are specialized to cases of 2-dimensional ray systems and further reduced to the special case of depth dependent sound speed commonly encountered in underwater acoustics.

There is more than one motive for this work. First is that these equations are useful for practical calculations in acoustics and the author wants to make them readily available. Second is to provide a pedagogical presentation of the use of differential geometry, and more specifically the geometry of null vectors, for deriving the Jacobi equation. Last is to impress upon the reader the full power of conformal equivalence and the form invariant nature of the paraxial system when viewed from the paradigm of differential geometry, this is not present in other treatments. There are some subtleties involved and the reader should be aware that this paper will read more like a section from a text book or a set of lecture notes rather than a research article. The intent is to show enough steps to make derivations apparent without being pedantic.

## 2. Null geodesics and geodesic deviation

In this section we present the equations describing null geodesics and geodesic deviation in general. The equations for geodesics, internal frame field and geodesic deviation for a manifold, $\mathcal{M}$ endowed with a metric $g_{\alpha\beta}$ in coordinates $\{x^\mu\}$ with affine parameter λ are given below.

$$\frac{d^2 x^\mu}{d\lambda^2} + \Gamma^\mu{}_{\alpha\beta} \frac{dx^\alpha}{d\lambda} \frac{dx^\beta}{d\lambda} = 0 \tag{1}$$

$$\frac{d\hat{e}_J{}^\mu}{d\lambda} + \Gamma^\mu{}_{\alpha\beta} \frac{dx^\alpha}{d\lambda} \hat{e}_J{}^\beta = 0 \tag{2}$$

$$\frac{d^2 Y_J}{d\lambda^2} + K_{JI} Y_I = 0 \tag{3}$$

$$\Gamma^\mu{}_{\alpha\beta} = \frac{1}{2} g^{\mu\nu} \{\partial_\alpha g_{\nu\beta} + \partial_\beta g_{\nu\alpha} - \partial_\nu g_{\alpha\beta}\} \tag{4}$$

$$R^\mu{}_{\nu\alpha\beta} = \partial_\alpha \Gamma^\mu{}_{\nu\beta} - \partial_\beta \Gamma^\mu{}_{\nu\alpha} + \Gamma^\mu{}_{\rho\alpha} \Gamma^\rho{}_{\nu\beta} - \Gamma^\mu{}_{\rho\beta} \Gamma^\rho{}_{\nu\alpha} \tag{5}$$

$$Y_J = g_{\alpha\beta} Y^\alpha \hat{e}_J{}^\beta \tag{6}$$

$$K_{JI} = \hat{e}_J{}^\beta \hat{e}_I{}^\beta \frac{dx^\alpha}{d\lambda} \frac{dx^\alpha}{d\lambda} R_{\mu\nu\alpha\beta} \tag{7}$$

Equations 1 through 3 describe the dynamic ray tracing system, equations 4 and 5 define the connection and curvature from derived from the metric and equations 6 and 7 express the relevant quantities in the local internal frame of reference. The reader is directed to the literature for a detailed description of all quantities [5]. The first derivative of coordinates defines the 4-dim velocity of the ray (its tangent vector), $T^\alpha = \dot{x}^\alpha$. This obeys the null constraint relative to the acoustic metric, $g(T,T) = 0$. The internal basis obeys the orthonormality relations relative to the metric, $g(T, \hat{e}_I) = 0$, $g(\hat{e}_J, \hat{e}_I) = \delta_{JI}$, for $I, J = 1,2$. There is a third (null) basis vector, $L^\alpha$, which obeys, $g(T, L) = -1$, $g(L, L) = 0$, $g(L, \hat{e}_I) = 0$. This is not needed for the paraxial system. Hawking provides a nice description of these basis vectors in reference [9]. Reference [5] gives a description of an auxiliary bases $\tilde{e}_I$ that is purely spatial, *i.e.* does not contain a time component, and is related to the 2-form $\Lambda_I^{\alpha\beta} = T^\alpha \hat{e}_I^\beta - T^\beta \hat{e}_I^\alpha$. This will be used in the next section. Based on the above statements it turns out that the space part of the internal basis vectors will not always be unit vectors relative to the dot product (Euclidean metric) in 3-dim. We define a unit vector in 3-dim such that $\tilde{e}_I = \|\tilde{e}_I\| \hat{u}_I$. The absence of time dependence in $c$ means that there is a conserved quantity, $g_{00} dt/d\lambda + g_{0k} dx^k/d\lambda = \kappa_0$. For the cases presented here the metric is diagonal and we have $g_{00} dt/d\lambda = \kappa_0$. We absorb the constant in the definition of $\lambda$, which we are free to do since affine parameters are an equivalence class related by $a\lambda + b$, $a, b$ constant. A conformal transformation is defined by $\{g_{\alpha\beta}, d\lambda\} \to \{f(x^\mu) g_{\alpha\beta}, f(x^\mu) d\lambda\}$, where $f$ is a positive smooth function of coordinates.

The acoustic metric arises from a treatment of the method of characteristics applied to the equations of fluid mechanics. The form of the equation appearing in the acoustics literature, White [3], is given below.

$$g_{\alpha\beta} = \begin{bmatrix} -c^2 + v^2 & -v_i \\ -v_j & \delta_{ji} \end{bmatrix} \tag{8}$$

Other versions of this can be found in [2]. Equation 8 was the starting point of the author's work on paraxial systems [5], [6]. For this paper we specialize to $v_j = 0$ and reiterate that $\rho_0 =$ constant.

## 3. Derivation of the paraxial system in three representations

This section provides detailed derivations of the paraxial ray system presented in section 2. Steps are shown and justifications given. We present three different representations of the full paraxial ray equations.

We begin by noting the special assumptions regarding the form of the metric tensor; (1) that only the sound speed is included and (2) that the sound speed is independent of time. We proceed by looking at the various terms that arose from equation 4 of section 2 under these assumptions. A consequence of assumption (1) is that the metric is diagonal. We label coordinates $(t, x, y, z)$ as $(x^0, x^k)$ in tensors and connection coefficients.

$$\Gamma^0{}_{00} = \frac{1}{2}g^{00}\{\partial_0 g_{00} + \partial_0 g_{00} - \partial_0 g_{00}\} + \frac{1}{2}g^{0k}\{\partial_0 g_{k0} + \partial_0 g_{k0} - \partial_k g_{00}\} = 0$$

$$\Gamma^0{}_{0k} = \frac{1}{2}g^{00}\{\partial_0 g_{0k} + \partial_k g_{00} - \partial_0 g_{0k}\} = \frac{1}{2}g^{00}\partial_k g_{00}$$

$$\Gamma^k{}_{00} = \frac{1}{2}g^{kj}\{\partial_0 g_{j0} + \partial_0 g_{j0} - \partial_j g_{00}\} = -\frac{1}{2}g^{kj}\partial_j g_{00}$$

$$\Gamma^0{}_{ij} = \frac{1}{2}g^{00}\{\partial_j g_{0i} + \partial_i g_{0j} - \partial_0 g_{ij}\} = 0$$

$$\Gamma^k{}_{0j} = \frac{1}{2}g^{ki}\{\partial_0 g_{ij} + \partial_j g_{i0} - \partial_i g_{0j}\} = 0$$

$$\Gamma^0{}_{0k} = \frac{1}{2}g^{00}\{\partial_0 g_{0k} + \partial_k g_{00} - \partial_0 g_{0k}\} = \frac{1}{2}g^{00}\partial_k g_{00}$$

$$\Gamma^i{}_{jk} = \frac{1}{2}g^{im}\{\partial_j g_{mk} + \partial_k g_{mj} - \partial_m g_{jk}\}$$

The first connection is written out in long form to illustrate the proper meaning of contraction of indices. The entire second term vanishes due to the metric being diagonal, *i.e.* $g^{0k} = 0$, if fluid velocity were present this would not be the case. The first term vanishes because the metric components are independent of time. In the remaining steps we will not explicitly write out the off diagonal terms. We have only three classes of surviving connection coefficients under the assumptions given.

$$\Gamma^0{}_{0k} = \frac{1}{2}g^{00}\partial_k g_{00} \tag{9}$$

$$\Gamma^k{}_{00} = -\frac{1}{2} g^{kj} \partial_j g_{00} \tag{10}$$

$$\Gamma^i{}_{jk} = \frac{1}{2} g^{im}\{\partial_j g_{mk} + \partial_k g_{mj} - \partial_m g_{jk}\} \tag{11}$$

Also note that symmetry of the metric and Christoffel symbols is used occasionally so don't be surprised if index order is not what's expected. Symmetry of the metric is expressed as $g_{\alpha\beta} = g_{\beta\alpha}$. The Christoffel symbols obey the symmetry relation, $\Gamma^\mu{}_{\alpha\beta} = \Gamma^\mu{}_{\beta\alpha}$. Next are the Riemann curvature tensor components. Riemann obeys several symmetry relations which are helpful in reducing the amount of computation required for working out the tensors. Namely we have the following; $R_{\alpha\beta\mu\nu} = -R_{\alpha\beta\nu\mu}$, $R_{\alpha\beta\mu\nu} = -R_{\beta\alpha\mu\nu}$, $R_{\alpha\beta\mu\nu} = R_{\mu\nu\alpha\beta}$. In addition to these are two sets of identities called Bianchi identities which we will not require in this derivation. We split apart time and space indices as with the Christoffel symbols. For evaluating the sectional curvature in the internal frame field basis in equation 7 we will find it easier to work with Riemann having all indices lowered.

$$R_{\mu\nu\alpha\beta} = g_{\mu\sigma}(\partial_\alpha \Gamma^\sigma{}_{\nu\beta} - \partial_\beta \Gamma^\sigma{}_{\nu\alpha} + \Gamma^\sigma{}_{\rho\alpha}\Gamma^\rho{}_{\nu\beta} - \Gamma^\sigma{}_{\rho\beta}\Gamma^\rho{}_{\nu\alpha}) \tag{12}$$

Based on symmetry there are only three classes of tensor components, those with two time and two space indices, those with three space indices and finally those with all four space indices.

$$R_{0k0j} = g_{00}(\partial_0 \Gamma^0{}_{kj} - \partial_j \Gamma^0{}_{k0} + \Gamma^0{}_{i0}\Gamma^i{}_{kj} - \Gamma^0{}_{0j}\Gamma^0{}_{k0})$$
$$= g_{00}(-\partial_j \Gamma^0{}_{k0} + \Gamma^0{}_{i0}\Gamma^i{}_{kj} - \Gamma^0{}_{0j}\Gamma^0{}_{k0})$$

$$R_{0kij} = g_{00}(\partial_i \Gamma^0{}_{kj} - \partial_j \Gamma^0{}_{ki} + \Gamma^0{}_{0i}\Gamma^0{}_{kj} - \Gamma^0{}_{0j}\Gamma^0{}_{ki}) = 0$$

$$R_{kijm} = g_{kn}(\partial_j \Gamma^n{}_{im} - \partial_m \Gamma^n{}_{ij} + \Gamma^n{}_{lj}\Gamma^l{}_{im} - \Gamma^n{}_{lm}\Gamma^l{}_{ij})$$

Of the three classes of tensor components only two classes survive. All others may be implied by the symmetry relations of Riemann. The reader is urged to remember that this is only valid for the assumptions we are working with and not a reduction to expect in general. For future reference the important terms are given below.

$$R_{0k0j} = g_{00}(-\partial_j \Gamma^0{}_{k0} + \Gamma^0{}_{i0}\Gamma^i{}_{kj} - \Gamma^0{}_{0j}\Gamma^0{}_{k0}) \tag{13}$$

$$R_{kijm} = g_{kn}(\partial_j \Gamma^n{}_{im} - \partial_m \Gamma^n{}_{ij} + \Gamma^n{}_{lj}\Gamma^l{}_{im} - \Gamma^n{}_{lm}\Gamma^l{}_{ij}) \tag{14}$$

We are now in a position to work out the ray equation and geometric spread equation for each of three cases of interest.

## Case 1: Arc length parameterization

In case 1 we have the metric $-cdt^2 + c^{-1}d\vec{r} \cdot d\vec{r}$. In matrix form this reads,

$$g_{\alpha\beta} = \begin{bmatrix} -c & 0 \\ 0 & c^{-1}[id_{3\times 3}] \end{bmatrix} \qquad g^{\alpha\beta} = \begin{bmatrix} -c^{-1} & 0 \\ 0 & c[id_{3\times 3}] \end{bmatrix}.$$

The Christoffel symbols for this metric are as follows.

$$\Gamma^0{}_{0k} = \frac{1}{2}\frac{\partial_k c}{c}$$

$$\Gamma^k{}_{00} = \frac{1}{2}c\partial_k c$$

$$\Gamma^i{}_{jk} = -\frac{1}{2}\{\delta_{ik}\delta_{mj} + \delta_{ij}\delta_{mk} - \delta_{jk}\delta_{mi}\}\frac{\partial_m c}{c}$$

From this we write the geodesic equations.

$$\frac{d^2 t}{d\lambda^2} + 2\Gamma^0{}_{0k}\frac{dt}{d\lambda}\frac{dx^k}{d\lambda} = 0$$

$$\frac{d^2 x^k}{d\lambda^2} + \Gamma^k{}_{00}\left(\frac{dt}{d\lambda}\right)^2 + \Gamma^k{}_{ij}\frac{dx^i}{d\lambda}\frac{dx^j}{d\lambda} = 0$$

Substituting the explicit expressions for the Christoffel symbols yields the following.

$$\frac{d^2 t}{d\lambda^2} + \frac{dt}{d\lambda}\left(\frac{d\vec{r}}{d\lambda}\cdot \vec{\nabla}\ln c\right) = 0$$

$$\frac{d^2 x^k}{d\lambda^2} + \frac{1}{2}c\partial_k c\left(\frac{dt}{d\lambda}\right)^2 - \frac{1}{2}\{\delta_{ik}\delta_{mj} + \delta_{kj}\delta_{mi} - \delta_{ji}\delta_{mk}\}\frac{\partial_m c}{c}\frac{dx^i}{d\lambda}\frac{dx^j}{d\lambda} = 0$$

The equation for time was easy to express in vector form. The dot is the ordinary inner product on Euclidean space. The equation for space coordinates along the geodesic is reduced further by use of the identity $\|d\vec{r}/dt\| = c$, and the fact that time is a monotonically increasing function of affine parameter. The last statement means that we can formally invert this relation to get $\lambda(t)$ and write $d\lambda/dt = 1/(dt/d\lambda)$. Distributing terms through the last set of brackets leads to the following form of the equation in vector notation.

$$\frac{d^2 \vec{r}}{d\lambda^2} + \frac{1}{2}c\vec{\nabla}c\left(\frac{dt}{d\lambda}\right)^2 - \frac{d\vec{r}}{d\lambda}\left(\frac{d\vec{r}}{d\lambda}\cdot \vec{\nabla}\ln c\right) + \frac{1}{2}\left\|\frac{d\vec{r}}{d\lambda}\right\|^2 \frac{\vec{\nabla}c}{c} = 0$$

So far no substitutions have been made. Here is where the above comments come in handy. We combine the second and fourth terms and factor.

$$\frac{d^2\vec{r}}{d\lambda^2} + \frac{1}{2}\left(c^2\left(\frac{dt}{d\lambda}\right)^2 + \left\|\frac{d\vec{r}}{d\lambda}\right\|^2\right)\frac{\vec{\nabla}c}{c} - \frac{d\vec{r}}{d\lambda}\left(\frac{d\vec{r}}{d\lambda}\cdot\vec{\nabla}\ln c\right)$$
$$= \frac{d^2\vec{r}}{d\lambda^2} + \frac{1}{2}\left(c^2 + \left\|\frac{d\vec{r}}{dt}\right\|^2\right)\left(\frac{dt}{d\lambda}\right)^2\frac{\vec{\nabla}c}{c} - \frac{d\vec{r}}{d\lambda}\left(\frac{d\vec{r}}{d\lambda}\cdot\vec{\nabla}\ln c\right)$$
$$= \frac{d^2\vec{r}}{d\lambda^2} + \frac{1}{2}(c^2 + c^2)\left(\frac{dt}{d\lambda}\right)^2\frac{\vec{\nabla}c}{c} - \frac{d\vec{r}}{d\lambda}\left(\frac{d\vec{r}}{d\lambda}\cdot\vec{\nabla}\ln c\right)$$
$$= \frac{d^2\vec{r}}{d\lambda^2} + \left(\frac{dt}{d\lambda}\right)^2 c\vec{\nabla}c - \frac{d\vec{r}}{d\lambda}\left(\frac{d\vec{r}}{d\lambda}\cdot\vec{\nabla}\ln c\right)$$

The first equality comes from application of $d/d\lambda = (dt/d\lambda)d/dt$. The second equality comes from applying the null constraint, which is nothing more than stating that the ray speed is the local sound speed. The rest is just cleaning up. In this representation the tangent vector for null geodesics is $\dot{x}^\alpha = (c^{-1}, \hat{n})$ which can be used to eliminate $(dt/d\lambda)$. The final form of our equation is,

$$\frac{d^2\vec{r}}{d\lambda^2} + \frac{1}{c}\vec{\nabla}c - \frac{d\vec{r}}{d\lambda}\left(\frac{d\vec{r}}{d\lambda}\cdot\vec{\nabla}\ln c\right) = 0. \tag{15}$$

Next we derive the sectional curvature. For this we need the basis vectors for the internal frame field. The spatial part of the 4-dim tangent vector is the wavefront normal. The basis vectors contain a time component so we appeal to the auxiliary basis described in appendix C of [5] which has a more physical interpretation. The auxiliary vector in this representation is $\tilde{e}_I^\alpha = (0, \sqrt{c}\hat{u}_I)$, is always purely spatial and the unit vectors $\hat{u}_I$ obey the orthonormality conditions, $\hat{u}_I \cdot \hat{n} = 0$, $\hat{u}_I \cdot \hat{u}_J = \delta_{IJ}$, relative to the dot product. The Riemann curvature tensor components in this representation are,

$$R_{0i0j} = \frac{1}{2c}\left(c\partial_i\partial_j c - \partial_i c \partial_j c\right),$$

$$R_{kijl} = \frac{1}{2c^3}\left(\delta_{kj}c\partial_i\partial_l c - \delta_{ij}c\partial_k\partial_l c - \delta_{kl}c\partial_i\partial_j c + \delta_{il}c\partial_k\partial_j c\right)$$
$$- \frac{1}{2c^3}\left(\delta_{kj}\partial_i c\partial_l c - \delta_{ij}\partial_k c\partial_l c - \delta_{kl}\partial_i c\partial_j c + \delta_{il}\partial_k c\partial_j c\right)$$
$$+ \frac{1}{4c^3}\left(\delta_{kl}\delta_{ij} - \delta_{kj}\delta_{il}\right)\vec{\nabla}c \cdot \vec{\nabla}c.$$

To calculate the sectional curvature we need to project Riemann into local geodesic frame field.

$$K_{IJ} = \frac{1}{4}R_{\mu\nu\alpha\beta}\Lambda_I^{\mu\nu}\Lambda_J^{\alpha\beta} = R_{0i0j}\Lambda_I^{0i}\Lambda_J^{0j} + \frac{1}{4}R_{kijl}\Lambda_I^{ki}\Lambda_J^{jl}$$

The auxiliary basis is contained in the 2-form $\Lambda_I^{\mu\nu}$, whose components are $\Lambda_J^{0j} = \dot{x}^0 \tilde{e}_J^j$ and $\Lambda_I^{ki} = \dot{x}^k \tilde{e}_I^i - \dot{x}^i \tilde{e}_I^k$. The first term is straight forward to work out.

$$R_{0i0j}\Lambda_I^{0i}\Lambda_J^{0j} = \frac{1}{2c}(c\partial_i\partial_j c - \partial_i c \partial_j c)\frac{1}{c^2}c\hat{u}_I^i\hat{u}_J^j = \frac{1}{2c^2}\left(c\hat{u}_I^i\hat{u}_J^j\partial_i\partial_j c - (\hat{u}_I \cdot \vec{\nabla}c)(\hat{u}_J \cdot \vec{\nabla}c)\right)$$

To work out the second term we note that the space components of the 2-form can be expressed in terms of the dual of the cross product of the basis vector with the wavefront normal. Specifically we define a convention such that $\hat{n} \times \hat{u}_1 = \hat{u}_2$, $\hat{u}_2 \times \hat{n} = \hat{u}_1$, $\hat{u}_1 \times \hat{u}_2 = \hat{n}$. Hence we may write,

$$\Lambda_I^{ki} = \sqrt{c}(n^k u_I^i - n^i u_I^k) = \sqrt{c}\varepsilon^{kim}u_J^m \epsilon_{IJ},$$

where we have introduced the three dimensional Levi-Civita symbol $\varepsilon^{kim}$ to account for the behavior of the cross product and the two dimensional $\epsilon_{IJ}$, $\epsilon_{11} = \epsilon_{22} = 0$, $\epsilon_{12} = -\epsilon_{21} = 1$. The second term may now be reduced as follows.

$$\begin{aligned}\frac{1}{4}R_{kijl}\Lambda_I^{ki}\Lambda_J^{jl} &= \frac{1}{4}\Bigg(\frac{1}{2c^2}4(\delta^{il}\delta^{mn} - \delta^{im}\delta^{ln})c\partial_i\partial_l c - \frac{1}{2c^2}4(\delta^{il}\delta^{mn} - \delta^{im}\delta^{ln})\partial_i c\partial_l c \\&\quad -\frac{1}{2c^2}2\delta^{mn}\vec{\nabla}c \cdot \vec{\nabla}c\Bigg)u_K^m u_L^n \epsilon_{IK}\epsilon_{JL} \\&= \frac{1}{4}\Bigg(\frac{1}{2c^2}(4\delta_{IJ}c\nabla^2 c - 4\epsilon_{IK}\epsilon_{JL}u_K^m u_L^n c\partial_m\partial_n c) \\&\quad -\frac{1}{4c^2}\left(4\delta_{IJ}\vec{\nabla}c \cdot \vec{\nabla}c - 4\epsilon_{IK}\epsilon_{JL}(\hat{u}_K \cdot \vec{\nabla}c)(\hat{u}_L \cdot \vec{\nabla}c)\right) - \frac{1}{c^2}\delta_{IJ}\vec{\nabla}c \cdot \vec{\nabla}c\Bigg)\end{aligned}$$

Adding the two terms leads to the following expression for the sectional curvature.

$$\begin{aligned}K_{IJ} &= \frac{1}{2c^2}\left(c\hat{u}_I^i\hat{u}_J^j\partial_i\partial_j c - (\hat{u}_I \cdot \vec{\nabla}c)(\hat{u}_J \cdot \vec{\nabla}c)\right) + \frac{1}{2c^2}(\delta_{IJ}c\nabla^2 c - \epsilon_{IK}\epsilon_{JL}u_K^m u_L^n c\partial_m\partial_n c) \\&\quad - \frac{1}{2c^2}\delta_{IJ}\vec{\nabla}c \cdot \vec{\nabla}c + \frac{1}{4c^2}\epsilon_{IK}\epsilon_{JL}(\hat{u}_K \cdot \vec{\nabla}c)(\hat{u}_L \cdot \vec{\nabla}c) \\&= \frac{1}{2c^2}\bigg((\delta_{IK}\delta_{JL} - \epsilon_{IK}\epsilon_{JL})u_K^m u_L^n c\partial_m\partial_n c + \delta_{IJ}(c\nabla^2 c - \vec{\nabla}c \cdot \vec{\nabla}c) \\&\quad - \left(\delta_{IK}\delta_{JL} - \frac{1}{2}\epsilon_{IK}\epsilon_{JL}\right)(\hat{u}_K \cdot \vec{\nabla}c)(\hat{u}_L \cdot \vec{\nabla}c)\bigg)\end{aligned}$$

At this point we will not attempt to reduce this expression further. It is a little foreboding but we'll see later that is simplifies quite a bit for 2-dim ray systems and 1-dim layered media. For now we state the results for later reference.

$$\frac{d^2 Y_J}{ds^2} + K_{IJ}Y_I = 0 \tag{16}$$

$$K_{IJ} = \frac{1}{2c^2}\Big((\delta_{IK}\delta_{JL} - \epsilon_{IK}\epsilon_{JL})u_K^m u_L^n c\partial_m\partial_n c + \delta_{IJ}(c\nabla^2 c - \vec{\nabla}c\cdot\vec{\nabla}c)$$
$$- \Big(\delta_{IK}\delta_{JL} - \frac{1}{2}\epsilon_{IK}\epsilon_{JL}\Big)(\hat{u}_K\cdot\vec{\nabla}c)(\hat{u}_L\cdot\vec{\nabla}c)\Big) \tag{17}$$

One final point is the propagation of the basis vectors. This is needed to evaluate the Jacobi equation at each step of the procedure. This is given by the parallel transport equation, equation 2 of section 2.

$$\frac{d\hat{e}_I^0}{d\lambda} + \Gamma^0{}_{0k}\frac{dt}{d\lambda}\hat{e}_I^k + \Gamma^0{}_{0k}\frac{dx^k}{d\lambda}\hat{e}_I^0 = 0$$

$$\frac{d\hat{e}_I^k}{d\lambda} + \Gamma^k{}_{00}\frac{dt}{d\lambda}\hat{e}_I^0 + \Gamma^k{}_{ij}\frac{dx^i}{d\lambda}\hat{e}_I^k = 0$$

Using the Christoffel symbols for this metric leads to,

$$\frac{d\hat{e}_I^0}{d\lambda} + \frac{1}{2}\frac{\partial_k c}{c}\frac{dt}{d\lambda}\hat{e}_I^k + \frac{1}{2}\frac{\partial_k c}{c}\frac{dx^k}{d\lambda}\hat{e}_I^0 = \frac{d\hat{e}_I^0}{d\lambda} + \frac{1}{2c^2}\hat{e}_I\cdot\vec{\nabla}c + \frac{1}{2c}\hat{e}_I^0(\hat{n}\cdot\vec{\nabla}c)$$
$$= \frac{d\hat{e}_I^0}{d\lambda} + \frac{1}{2c^2}\hat{e}_I\cdot\vec{\nabla}c + \frac{1}{2c^2}(\hat{e}_I\cdot\hat{n})(\hat{n}\cdot\vec{\nabla}c),$$

$$\frac{d\hat{e}_I^k}{d\lambda} + \frac{1}{2}c\partial_k c\frac{dt}{d\lambda}\hat{e}_I^0 - \frac{1}{2c}\{\delta_{ik}\partial_j c + \delta_{kj}\partial_i c - \delta_{ji}\partial_k c\}\hat{n}^i\hat{e}_I^j$$
$$= \frac{d\hat{e}_I^k}{d\lambda} + \frac{1}{2}c\partial_k c\frac{dt}{d\lambda}\hat{e}_I^0 - \frac{1}{2c}\{\hat{n}^k(\hat{e}_I\cdot\vec{\nabla}c) + \hat{e}_I^k(\hat{n}\cdot\vec{\nabla}c) - (\hat{e}_I\cdot\hat{n})\partial_k c\}$$
$$= \frac{d\hat{e}_I^k}{d\lambda} + \frac{1}{2c}\partial_k c(\hat{e}_I\cdot\hat{n}) - \frac{1}{2c}\{\hat{n}^k(\hat{e}_I\cdot\vec{\nabla}c) + \hat{e}_I^k(\hat{n}\cdot\vec{\nabla}c) - (\hat{e}_I\cdot\hat{n})\partial_k c\}.$$

We have made use of the constraints on the basis vectors to replace certain quantities. For example the constraint $g(\dot{x}, \hat{e}_I) = 0$ implies $\hat{e}_I^0 = \hat{n}\cdot\hat{e}_I/c$. This fact implies that the equation for the time component is redundant and can be used to reduce the space equation. In vector form we have,

$$\frac{d\hat{e}_I}{d\lambda} + \frac{1}{c}\vec{\nabla}c(\hat{e}_I\cdot\hat{n}) - \frac{1}{2c}\{\hat{n}(\hat{e}_I\cdot\vec{\nabla}c) + \hat{e}_I(\hat{n}\cdot\vec{\nabla}c)\} = 0, \tag{18}$$

$$\frac{d\hat{e}_I^0}{d\lambda} + \frac{1}{2c^2}\hat{e}_I\cdot\vec{\nabla}c + \frac{1}{2c^2}(\hat{e}_I\cdot\hat{n})(\hat{n}\cdot\vec{\nabla}c). \tag{19}$$

At this point we will not simplify things further. The astute reader may have seen several places along the way where substitutions could be made. We'll discuss this later with respect to all representations. Our initial goal is to provide a complete set of paraxial equations for a common set of variables in each of three representations.

*Case 2: Affine representation*

In case 2 our metric is $-c^2 dt^2 + d\vec{r} \cdot d\vec{r}$, or in matrix form given below.

$$g_{\alpha\beta} = \begin{bmatrix} -c^2 & 0 \\ 0 & id_{3\times 3} \end{bmatrix} \quad g^{\alpha\beta} = \begin{bmatrix} -c^{-2} & 0 \\ 0 & id_{3\times 3} \end{bmatrix}$$

One could work out the connection coefficients by hand using either the definition or by applying a conformal transformation. Either way is about the same amount of work for metrics of this form. In this case we have only two surviving elements to the connection.

$$\Gamma^0{}_{0k} = \frac{\partial_k c}{c}$$

$$\Gamma^k{}_{00} = c\partial_k c$$

The geodesic equation in this representation is much simpler to work out. Since so much effort went into providing steps for this first case we will present the results here for case 2 without showing as many steps. In this representation the null tangent vector for the geodesic is $\dot{x}^\alpha = (c^{-2}, c^{-1}\hat{n})$, the spatial component represents the slowness vector.

$$\frac{d^2 t}{d\lambda^2} + 2\frac{1}{c^3}\left(\frac{d\vec{r}}{d\lambda} \cdot \vec{\nabla} c\right) = 0 \tag{20}$$

$$\frac{d^2 \vec{r}}{d\lambda^2} + \frac{1}{c^3}\vec{\nabla} c = 0 \tag{21}$$

In this representation the basis vectors contain time components but we don't need to use these. Based on the nature of the surviving components of the connection we only have to deal with one class of non-vanishing Riemann tensor components, $R_{0k0j}$.

$$R_{0k0j} = g_{00}\left(-\partial_j \Gamma^0{}_{k0} - \Gamma^0{}_{0j}\Gamma^0{}_{k0}\right) = c^2\left(\partial_j(c^{-1}\partial_k c) + c^{-2}\partial_j c \partial_k c\right) = c\partial_j \partial_k c$$

To calculate the sectional curvature evaluated in the local internal frame field use the auxiliary basis, $\tilde{e}_I$.

$$K_{IJ} = \frac{1}{4}R_{\mu\nu\alpha\beta}\Lambda_I^{\mu\nu}\Lambda_J^{\alpha\beta} = \frac{1}{4}\left(R_{0i0j}\Lambda_I^{0i}\Lambda_J^{0j} + R_{i00j}\Lambda_I^{i0}\Lambda_J^{0j} + R_{0ij0}\Lambda_I^{0i}\Lambda_J^{j0} + R_{i0j0}\Lambda_I^{i0}\Lambda_J^{j0}\right)$$

$$= R_{0i0j}\Lambda_I^{0i}\Lambda_J^{0j} = (\dot{x}^0)^2 \tilde{e}_I^i \tilde{e}_J^j c\partial_j\partial_k c = \tilde{e}_I^i \tilde{e}_J^j \frac{\partial_j \partial_k c}{c^3}$$

Of all the cases we present here this is probably the easiest representation in which to express the paraxial ray system. There are very few terms to the connection and Riemann tensor and a fairly simple interpretation to the curvature tensor. Treating the internal basis as a rotation at a fixed point along the ray path in 4-dim the sectional curvature is a 2-by-2 matrix of partial derivatives in the space orthogonal to the direction of the wavefront normal (or ray path direction in this case). The Jacobi equation (a.k.a. the geodesic deviation equation) is,

$$\frac{d^2 Y_J}{d\lambda^2} + \{\tilde{e}_I^i \tilde{e}_J^j c^{-3} \partial_j \partial_k c\} Y_I = 0 \tag{23}$$

As in the last case we provide the equations to propagate the basis vectors along the geodesic. Direct substitution leads to the following.

$$\frac{d\hat{e}_I^0}{d\lambda} + \frac{1}{c}\left(\hat{e}_I^0 \frac{d\vec{x}}{d\lambda} \cdot \vec{\nabla} c + \frac{dt}{d\lambda} \vec{e}_I \cdot \vec{\nabla} c\right) = 0$$

$$\frac{d\vec{e}_I}{d\lambda} + \frac{dt}{d\lambda} \hat{e}_I^0 c \vec{\nabla} c = 0$$

After a little substituting we arrive at the following.

$$\frac{d\hat{e}_I^0}{d\lambda} + \frac{1}{c^3}\left((\vec{e}_I \cdot \hat{n})(\hat{n} \cdot \vec{\nabla} c) + \vec{e}_I \cdot \vec{\nabla} c\right) = 0 \tag{24}$$

$$\frac{d\vec{e}_I}{d\lambda} + \frac{1}{c^2}(\vec{e}_I \cdot \hat{n})\vec{\nabla} c = 0 \tag{25}$$

*Case 3: Time parameterization*

Next we move on to case 3. Here the metric is $-dt^2 + c^{-2} d\vec{r} \cdot d\vec{r}$. The matrix form is provided below.

$$g_{\alpha\beta} = \begin{bmatrix} -1 & 0 \\ 0 & c^{-2}[id_{3\times 3}] \end{bmatrix} \qquad g^{\alpha\beta} = \begin{bmatrix} -1 & 0 \\ 0 & c^{2}[id_{3\times 3}] \end{bmatrix}$$

In this case there is only one surviving class of connection coefficients, namely those with three space indices.

$$\Gamma^i{}_{jk} = -\{\delta_{ik}\delta_{mj} + \delta_{ij}\delta_{mk} - \delta_{jk}\delta_{mi}\}\frac{\partial_m c}{c}$$

Now the geodesic equation for time takes the following form.

$$\frac{d^2 t}{d\lambda^2} = 0 \tag{26}$$

This can be solved by inspection to give, $t = a\lambda + b$. In other words in this representation time is a member of the equivalence class of affine parameters. The equation for the space coordinates is

$$\frac{d^2 x^k}{d\lambda^2} - \{\delta_{jk}\delta_{mi} + \delta_{ki}\delta_{mj} - \delta_{ji}\delta_{mk}\}\frac{\partial_m c}{c}\frac{dx^i}{d\lambda}\frac{dx^j}{d\lambda}$$

$$= \frac{d^2 \vec{r}}{d\lambda^2} - 2\frac{d\vec{r}}{d\lambda}\left(\frac{d\vec{r}}{d\lambda}\cdot \vec{\nabla}\ln c\right) + \left\|\frac{d\vec{r}}{d\lambda}\right\|^2 \frac{1}{c}\vec{\nabla}c$$

$$= a^2\left(\frac{d^2\vec{r}}{dt^2} - 2\frac{d\vec{r}}{dt}\left(\frac{d\vec{r}}{dt}\cdot\vec{\nabla}\ln c\right) + \left\|\frac{d\vec{r}}{dt}\right\|^2\frac{1}{c}\vec{\nabla}c\right)$$

$$= a^2\left(\frac{d^2\vec{r}}{dt^2} - 2\frac{d\vec{r}}{dt}\left(\frac{d\vec{r}}{dt}\cdot\vec{\nabla}\ln c\right) + c\vec{\nabla}c\right)$$

The first equality is just expressing summation notation as vectors using the ordinary dot product on Euclidean space. The second equality makes use of the fact that time is an affine parameter (*i.e.* the explicit solution to the time equation) and the last equality makes use of the same null constraint as used in case 1. The final form for the ray equation in this case is given below.

$$\frac{d^2\vec{r}}{dt^2} - 2\frac{d\vec{r}}{dt}\left(\frac{d\vec{r}}{dt}\cdot\vec{\nabla}\ln c\right) + c\vec{\nabla}c = 0 \tag{27}$$

The curvature tensor components are now calculated for this case. There is only one surviving class of curvature tensor components, those containing 4 space indices, and there is quite a bit of serendipitous cancellation of terms. Nothing is gained by writing this out long hand and the reader can verify the following using the connection coefficients.

$$R_{nijm} = c^{-3}(\delta_{im}\partial_n\partial_j c + \delta_{nj}\partial_i\partial_m c - \delta_{nm}\partial_i\partial_j c - \delta_{ij}\partial_n\partial_m c) + c^{-4}(\delta_{nm}\delta_{ij} - \delta_{nj}\delta_{im})\vec{\nabla}c\cdot\vec{\nabla}c$$

For the sectional curvature we need the internal basis vectors. In this case things simplify a little. Since time is an affine parameter in this case and there are no connection coefficients with time indices the equation for the evolution of $\hat{e}_I^0$ is trivial, $d\hat{e}_I^0/dt = 0$. We choose the initial value of the basis vectors to be purely spatial, and they remain so in this representation. As a result of this the auxiliary basis satisfies $\tilde{e}_I = \hat{e}_I$ for all time. In affine representation the auxiliary basis was automatically a unit vector due to the form of the metric tensor. The conformal transformation will affect the basis vectors as we must maintain orthonormality, $g(\dot{x},\dot{x}) = 0$, $g(\dot{x},\hat{e}_I) = 0$, $g(\hat{e}_J,\hat{e}_I) = \delta_{IJ}$. To accomplish this the basis vectors are scaled as follows, $\hat{e}_I = \tilde{e}_I = c\hat{u}_I$, where $\hat{u}_I \cdot \hat{u}_J = \delta_{IJ}$ relative to the ordinary dot product (a.k.a. the Euclidian metric tensor). With these points in place the sectional curvature is,

$$K_{IJ} = R_{lijm}\dot{x}^l\dot{x}^j\tilde{e}_I^i\tilde{e}_J^m$$
$$= c^{-3}(c^4(\hat{u}_I \cdot \hat{u}_J)\hat{n}^i\hat{n}^j\partial_l\partial_j c + c^4\hat{u}_I^i\hat{u}_J^m\partial_i\partial_m c) + c^{-4}\left(-c^4(\hat{u}_I \cdot \hat{u}_J)\right)\vec{\nabla}c \cdot \vec{\nabla}c$$
$$= (\delta_{IJ}\hat{n}^i\hat{n}^j + \hat{u}_I^i\hat{u}_J^j)c\partial_i\partial_j c - \delta_{IJ}\vec{\nabla}c\cdot\vec{\nabla}c$$

Though this is a little more complex that the previous case it is still compact. The Jacobi equation for this representation is given below.

$$\frac{d^2Y_J}{dt^2} + \{(\delta_{IJ}\hat{n}^i\hat{n}^j + \hat{u}_I^i\hat{u}_J^j)c\partial_i\partial_j c - \delta_{IJ}\vec{\nabla}c \cdot \vec{\nabla}c\}Y_I = 0 \tag{28}$$

The propagation equation for the basis vectors is.

$$\frac{d\tilde{e}_I}{d\lambda} - \left(\tilde{e}_I(\hat{n} \cdot \vec{\nabla}c) + \hat{n}(\tilde{e}_I \cdot \vec{\nabla}c)\right) = 0 \tag{29}$$

All of these results are summed up in table 1 below. Each row is a complete set of paraxial ray equations. The left column gives the affine parameter used, second column is the equation set, column three gives the conformal transformation required to get us to the next row in the table and the last column provides an equation number for future reference.

| Affine parameter | Ray equations | Conformal transform | Eq. Num. |
|---|---|---|---|
| s | $\ddot{t} + \dot{t}(\dot{\vec{r}} \cdot \vec{\nabla}\ln c) = 0$<br>$\ddot{\vec{r}} - \dot{\vec{r}}(\dot{\vec{r}} \cdot \vec{\nabla}\ln c) + \dot{t}^2 c\vec{\nabla}c = 0$<br>Basis = Equations 18, 19<br>Deviation = Equations 16, 17 | $ds = d\lambda/c$ | (30) |
| $\lambda$ | $\ddot{t} + 2\dot{t}(\dot{\vec{r}} \cdot \vec{\nabla}\ln c) = 0$<br>$\ddot{\vec{r}} - \dot{t}^2 c\vec{\nabla}c = 0$<br>$\dot{\tilde{e}}_I^0 + c^{-3}\left((\tilde{e}_I \cdot \hat{n})(\hat{n} \cdot \vec{\nabla}c) + \tilde{e}_I \cdot \vec{\nabla}c\right) = 0$<br>$\dot{\tilde{e}}_I + c^{-2}(\tilde{e}_I \cdot \hat{n})\vec{\nabla}c = 0$<br>$\ddot{Y}_J + \{\tilde{e}_I^i \tilde{e}_J^j c^{-3}\partial_j\partial_k c\}Y_I = 0$ | $d\lambda = c^2 dt$ | (31) |
| t | $\ddot{t} = 0$<br>$\ddot{\vec{r}} - 2\dot{\vec{r}}(\dot{\vec{r}} \cdot \vec{\nabla}\ln c) + c\vec{\nabla}c = 0$<br>$\dot{\tilde{e}}_I - \left(\tilde{e}_I(\hat{n} \cdot \vec{\nabla}c) + \hat{n}(\tilde{e}_I \cdot \vec{\nabla}c)\right) = 0$<br>$\ddot{Y}_J + \{(\delta_{IJ}\hat{n}^i\hat{n}^j + \hat{u}_I^i\hat{u}_J^j)c\partial_i\partial_j c - \delta_{IJ}\vec{\nabla}c \cdot \vec{\nabla}c\}Y_I = 0$ | $dt = ds/c$ | (32) |

**Table 1** – Paraxial ray system in three representations

The propagation equations provide the coordinates of the ray in space-time as well as the pseudo-orthonormal basis set $\hat{e}_I$, from which one can determine $\hat{n}$ and $\hat{u}_I$ form the constraints. In the discussion section we will introduce a simplification for this that determines $\hat{u}_I$ directly.

### *Reduction to 2-dim and 1-dim environmental profiles*

Things get simple pretty quick when we reduce the problem to one in which the local sound speed depends on only one or two spatial variables. There are two cases of interest that we now consider. The first is a purely depth dependent profile. This is the typical case studied in underwater acoustics, $c = c(z)$. Such cases do not exhibit "theta coupling", a term used in underwater acoustics and oceanography to refer to the fact that in general ray paths contain torsion and will twist out of their initial osculating plane. Here we can solve the ray system in one vertical plane and rotate these results by any azimuthal angle to get the full 3-dim solution to the ray system. These system are truly 2-dimensional and are sometimes referred to an N-by-2dim problems where N refers to the number of vertical planes used to fill space. We can choose any vertical plane without loss of generality and rotate coordinates such that this is the x-z plane in Cartesian space. In this vertical plane we can choose as a basis $\hat{n} = [n_x, 0, n_z]^T$, the wavefront normal and $\hat{u}_1 = [-n_z, 0, n_x]^T$. A complete treatment in the formalism of this paper would have a third basis vector $\hat{u}_2 = [0, 1, 0]^T$. For our purposes we ignore this basis vector. It is easy to see by inspection of the Riemann tensor that the sectional curvature in this direction, as well as the cross terms, will vanish leaving only the term $K_{11}$. The geometric spread in the horizontal plane will go as $1/\lambda$ for whatever affine parameter is being used.

When the local sound speed depends on only one variable there is a significant amount of symmetry in the system. The principle of isometry can be applied to the ray system to derive conserved currents related to each cyclic variable. The reader is referred to reference [8] for a detailed account of this but in a nut shell we arrive at the well-known depth dependent ray integrals as a solution to the ray system. This is seen as a result of there being enough symmetry to reduce the second order ODE system to a set of first order ODE's. These are equivalent to Snell's law and are expressed as $n_x/c = \text{constant} \equiv \alpha_x$ and $n_y/c = \text{constant} \equiv \alpha_y$, from which we define $\alpha = \sqrt{\alpha_x^2 + \alpha_y^2}$.

We are primarily interested in the form of the Jacobi equation in such cases. Applying these considerations to equation 19 gives us the following for the curvature.

$$\frac{1}{2c^2}\left(c\nabla^2 c - \vec{\nabla}c \cdot \vec{\nabla}c + \hat{u}^z \hat{u}^z cc'' - (\hat{u}^z c')^2\right) = \frac{1}{2c^2}(1 + (\hat{n}^x)^2)(cc'' - c'c')$$

$$= \frac{1}{2c^2}(1 + c^2\alpha^2)(cc'' - c'c')$$

In case 2 we arrive at $(\tilde{e}_1^z)^2 c^{-3}\partial_z^2 c = (n^x)^2 c^{-3}\partial_z^2 c = \alpha_x^2 c''/c$. Finally in case 3 we have $((\hat{n}^z)^2 + (\hat{u}_1^z)^2)c\partial_z^2 c - \vec{\nabla}c \cdot \vec{\nabla}c = ((\hat{n}^z)^2 + (\hat{n}^x)^2)\partial_z^2 c - \vec{\nabla}c \cdot \vec{\nabla}c = cc'' - c'^2$. All cases are simple but there is something rather nice about the time parameterization in that it does not depend on the launch angle of the ray, *i.e.* the initial conditions, and seems to relate all the focusing properties of the medium only to the medium.

The next case we consider is a 2-dim space with a sound speed depending on both coordinates, (x, z). This is more for academic interest but does have some practical use. Studies in ray chaos treat an approximately 2-dim system by considering rays traveling initially near the (x, z) plane and ignoring the twisting of the ray paths. Rays launched in the (x, z) plane will remain in that plane as they propagate so this subset of rays will make a true 2-dim system. Once again we use the basis vectors presented in the 1-dim case to reduce the curvature tensors when appropriate. Case 1 reduces to,

$$K_{11} = \frac{1}{2c^2}\left((\delta_{11}\delta_{11})u_1^m u_1^n c\partial_m\partial_n c + \delta_{IJ}(c\nabla^2 c - \vec{\nabla}c \cdot \vec{\nabla}c) - (\delta_{11}\delta_{11})(\hat{u}_1 \cdot \vec{\nabla}c)(\hat{u}_1 \cdot \vec{\nabla}c)\right)$$

$$= \frac{1}{2c^2}\left(u_1^m u_1^n c\partial_m\partial_n c - (\hat{u}_1 \cdot \vec{\nabla}c)(\hat{u}_1 \cdot \vec{\nabla}c) + (c\nabla^2 c - \vec{\nabla}c \cdot \vec{\nabla}c)\right).$$

All terms containing $\epsilon_{JM}$ vanish. In case 2 we already have a simple interpretation that survives. Defining a normal coordinate $\xi$, we can express the sectional curvature as $c^{-3}\partial_\xi^2 c$, which states that the curvature in the direction normal to the ray path governs the local focusing at that point. The third case reduces as follows,

$$
\begin{aligned}
(\hat{n}^i \hat{n}^j &+ \hat{u}_I^i \hat{u}_J^j) c \partial_i \partial_j c - \vec{\nabla} c \cdot \vec{\nabla} c \\
&= (\hat{n}^x \hat{n}^x + \hat{u}^x \hat{u}^x) c \partial_x \partial_x c + (\hat{n}^z \hat{n}^z + \hat{u}^z \hat{u}^z) c \partial_z \partial_z c + 2(\hat{n}^x \hat{n}^z + \hat{u}^x \hat{u}^z) c \partial_x \partial_z c \\
&\quad - \vec{\nabla} c \cdot \vec{\nabla} c \\
&= (\hat{n}^x \hat{n}^x + \hat{n}^z \hat{n}^z) c \partial_x \partial_x c + (\hat{n}^z \hat{n}^z + \hat{n}^x \hat{n}^x) c \partial_z \partial_z c + 2(\hat{n}^x \hat{n}^z - \hat{n}^x \hat{n}^z) c \partial_x \partial_z c \\
&\quad - \vec{\nabla} c \cdot \vec{\nabla} c = \hat{n}^x \hat{n}^x c (\partial_x \partial_x c + \partial_z \partial_z c) + \hat{n}^z \hat{n}^z c (\partial_x \partial_x c + \partial_z \partial_z c) - \vec{\nabla} c \cdot \vec{\nabla} c \\
&= (\hat{n}^x \hat{n}^x + \hat{n}^z \hat{n}^z) c (\partial_x \partial_x c + \partial_z \partial_z c) - \vec{\nabla} c \cdot \vec{\nabla} c = c \nabla^2 c - \vec{\nabla} c \cdot \vec{\nabla} c \,.
\end{aligned}
$$

Again it appears that in 2+1 dimensional with time parameterization the focusing properties of the medium do not depend on the particular ray path initial conditions. This is quite useful for evaluating the expectation of ray convergence or divergence before one even begins to start a ray trace and can be used to set step sizes and make other decisions regarding numerical procedures.

We present the curvature for each of our three cases and for each of three dimensions in table 2. In both of the special cases considered in this section, layered media and 2-dim acoustics, the auxiliary basis is superfluous since it can be determined by rotating the wave front normal by 90 degrees.

| Affine | Sectional Curvature $K_{IJ}$ | 2-dim curvature $c = c(x,z)$ | Depth dependent SSP $c = c(z)$ |
|---|---|---|---|
| s | Equation 19 | $\frac{1}{2c^2}\left(c\nabla^2 c - \vec{\nabla}c \cdot \vec{\nabla}c + \hat{u}^j \hat{u}^k c \partial_j \partial_k c - (\hat{u}^j \partial_j c)^2\right)$ | $\frac{(1+c^2\alpha^2)}{2c^2}(cc'' - c'c')$ |
| $\lambda$ | $\tilde{e}_I^i \tilde{e}_J^j c^{-3} \partial_j \partial_k c$ | $c^{-3} \partial_\xi^2 c$ | $\alpha^2 c''/c$ |
| t | $(\delta_{IJ}\hat{n}^i \hat{n}^j + \hat{u}_I^i \hat{u}_J^j) c \partial_i \partial_j c - \delta_{IJ} \vec{\nabla}c \cdot \vec{\nabla}c$ | $c\nabla^2 c - \vec{\nabla}c \cdot \vec{\nabla}c$ | $cc'' - c'^2$ |

**Table 2** – Curvature tensors for all three cases, reduced for 2-dim and depth dependent SSP

## 4. Discussion

Based on the above derivations a few comments are in order. Paraxial equations in various forms exist in the literature. Many are documented in chapter 4 of Cerveny [10]. The form of the paraxial equation varies quite a bit depending on the ray parameter used. Some parameterizations lead to forms of the equation with second and first order derivatives of the deviation vector and some even lead to inhomogeneous versions of the deviation equation. A point that cannot be stressed enough is that applying a parameter change to an ODE system is not the same as applying a conformal transformation to the paraxial system. One of the virtues of identifying the paraxial system with a null geodesic flow is that the ODE system describing paraxial rays is form invariant and there is a simple transformation law for the fields contained in the system [7]. Many of the treatments of paraxial ray systems and Gaussian beam systems derived from the Helmholtz equation effectively use a mixture of representations leading to increasing complexity. Many authors consider arc length to be the most natural place to start the development of the paraxial system. Based on the above discussion it is clear by inspection that the affine parameterization is by far the easiest to work with, even in 3-dimensions. The equations are not complex in form and simple to derive.

The unique nature of these types of systems makes them a little easier to work with. For example we note that in any of the representations one can deduce first order ODE's for the wavefront normal and the 3-dim unit vector, $\{\hat{n}, \hat{u}_I\}$. Since the curvatures have all been reduced to quadratic forms in $\hat{n}$ and $\hat{u}_I$ it makes sense to replace the equation for the pseudo-orthonormal basis with an equation for $\hat{u}_I$ and if possible include and equation for $\hat{n}$ and avoid any unnecessary algebra. The geodesic equation may also be manipulated to give an equation for the wavefront normal. This is easiest to do using the arc-length representation, equation 15, since we already have $d\vec{x}/d\lambda = d\vec{x}/ds = \hat{n}$.

$$\frac{d\hat{n}}{ds} + \frac{1}{c}\hat{n} \times (\vec{\nabla}c \times \hat{n}) = 0 \qquad (33)$$

We derive a similar equation for the unit vectors $\hat{u}_I$ using equation 29 in time representation.

$$\frac{d\hat{u}_I}{dt} - \hat{n}(\hat{u}_I \cdot \vec{\nabla}c) = 0 \qquad (34)$$

It is worth noting that these first order equation for the local orthonormal basis are enough to propagate the Jacobi equation but that full ray equation is need to locate where and when these values are to be attributed. By definition these unit vectors are invariant under a conformal transformation so one can express equations 33 and 34 in any representation by a simple change of variables, *i.e.* $ds = cdt$, *etc*. For the types of problems considered here we can get $\hat{n}$ and any step by the relation $\hat{n} = (d\vec{x}/d\lambda)/(dt/d\lambda)/c$. Furthermore a numerical implementation of the paraxial system would most likely be cast as a first order system by introducing new variables for the first derivatives, *i.e.* $p^\alpha \equiv dx^\alpha/d\lambda$, $P_I \equiv dY_I/d\lambda$. Hence we have the data we need to get $\hat{n}$ without much effort. Including equation 33 along with the others would likely be wasteful. The opposite is true for the basis vectors. In an arbitrary representation the equation for $\hat{u}_I$ requires all 4 components of $\hat{e}_I$ and the momentum for the geodesic. Replacing the basis equations with equation 34 would likely improve the procedure. Finally, the since $c$ is time independent the equation for time can be reduced to a first order ODE, $dt/d\lambda = \kappa_0/g_{00}$. For convenience table 1 is repeated here with these considerations in mind.

By inspection, and from the amount of work required to derive these quantities, the author's opinion is that the affine representation was the easiest to work in. There is a more physical interpretation of this parameter. Inspection of equation 21 illustrates that there is no dependence on the ray velocity. This form of the ray equation appears in the literature and the parameter referred to as Newton parameter to reflect the fact that the ray equation looks like Newton's second law of motion in a conservative force field. Defining $U = U_0 - c^{-2}/2$, we can write equation 21 as $d^2\vec{r}/d\lambda^2 = -\vec{\nabla}U$.

The author's use of the paraxial system implements the geodesic and geodesic deviation equations in their original abstract form, equations 1-7. This has been very useful and allows for immediate generalization to environments with non-zero background currents and explicit time dependence. Expressing the paraxial equations as we have done here helps one gain a deeper understanding of how the abstract equations relate to physical parameters. The three representations presented here are the most commonly used in acoustics but conformal symmetry allows for a continuum of other valid representations.

| Affine parameter | Ray equations | Conformal transform | Eq. Num. |
|---|---|---|---|
| $s$ Arc-length | $\dot{t} = c^{-1}$ <br> $\ddot{\vec{r}} - \dot{\vec{r}}(\dot{\vec{r}} \cdot \vec{\nabla} \ln c) + \dot{t}^2 c \vec{\nabla} c = 0$ <br> $\dot{\hat{u}}_I - \hat{n}(\hat{u}_I \cdot \vec{\nabla} \ln c) = 0$ <br> Deviation = Equations 16, 17 | $ds = d\lambda/c$ | (35) |
| $\lambda$ Newton | $\dot{t} = c^{-2}$ <br> $\ddot{\vec{r}} - \dot{t}^2 c \vec{\nabla} c = 0$ <br> $\dot{\hat{u}}_I - c^{-2}\hat{n}(\hat{u}_I \cdot \vec{\nabla} c) = 0$ <br> $\ddot{Y}_J + \{\hat{u}_I^i \hat{u}_J^j c^{-3} \partial_j \partial_k c\} Y_I = 0$ | $d\lambda = c^2 dt$ | (36) |
| $t$ Time | $\ddot{\vec{r}} - 2\dot{\vec{r}}(\dot{\vec{r}} \cdot \vec{\nabla} \ln c) + c\vec{\nabla} c = 0$ <br> $\dot{\hat{u}}_I - \hat{n}(\hat{u}_I \cdot \vec{\nabla} c) = 0$ <br> $\ddot{Y}_J + \{(\delta_{IJ}\hat{n}^i \hat{n}^j + \hat{u}_I^i \hat{u}_J^j) c \partial_l \partial_j c - \delta_{IJ} \vec{\nabla} c \cdot \vec{\nabla} c\} Y_I = 0$ | $dt = ds/c$ | (37) |

**Table 3** – Paraxial ray system in three representations

In this paper we have provided the complete paraxial ray equations appropriate for an environment described by a position dependent sound speed in three commonly used parameterizations. This was done using differential geometry. The purpose of this was to provide these to researchers who might find them useful as well as to provide a step by step derivation using concepts from differential geometry as a tutorial in these techniques. This may be of interest not only to acousticians but to students studying relativity or differential geometry as a case study. Connecting abstract differential geometry of Lorentzian manifolds to something as well studied as ray tracing can help bridge the gap from the abstract to the pragmatic use of these techniques. While this presentation focused on a special class of environments that were not too difficult to handle the author has used these techniques for acoustic modeling in the presence of 3-dim background winds and currents. Being able to relate the results of this paper to well-known results, such as those found in Cerveny [10], also helps build confidence in the technique.